\documentclass[12pt]{article}
\usepackage{a4}
\usepackage{amssymb}
\usepackage{graphicx}
\usepackage{epsf}
\usepackage{cite}
\newcommand{\be}{\begin{equation}}
\newcommand{\ee}{\end{equation}}
\newcommand{\bea}{\begin{eqnarray}}
\newcommand{\eea}{\end{eqnarray}}
\newcommand{\pt}{\partial}

\begin{document}

\begin{titlepage}
\title{Thermodynamic Geometry and Extremal Black Holes in String Theory}
\author{}
\date{Tapobrata Sarkar\thanks{\noindent tapo@iitk.ac.in}, 
Gautam Sengupta \thanks{\noindent sengupta@iitk.ac.in}, 
Bhupendra Nath Tiwari \thanks{\noindent bntiwari@iitk.ac.in}\\
\vspace{0.7 cm}
Department of Physics, \\ Indian Institute of Technology,\\
Kanpur 208016, India.}

\maketitle
\abstract{  We study a generalisation of thermodynamic geometry to 
degenerate quantum ground states at zero temperatures exemplified by 
charged extremal black holes in type II string theories. Several examples 
of extremal charged black holes with 
non degenerate thermodynamic geometries and finite but non zero 
state space scalar curvatures are established.
These include black 
holes described by D1-D5-P and D2-D6-NS5-P 
brane systems and also two charged small black holes in Type II string 
theories. We also explore the modifications to the state space geometry and 
the scalar curvature due to the higher derivative
contributions and string loop corrections as well as an exact 
entropy expression from quantum information theory.
Our construction describes state space geometries arising out of a 
possible limiting thermodynamic characterisation of degenerate 
quantum ground states at zero temperatures.}
\end{titlepage}

\section{Introduction.}\label{one}

Over the past decade, black hole thermodynamics has emerged as a crucial
theoretical laboratory to test issues of quantum gravity in the context of
string theories. The area has witnessed major advances especially toward a
resolution of the microscopic statistical basis underlying the macroscopic entropy
of extremal and near extremal black holes in string theory \cite{bhlex}.
Macroscopically black holes are known to be thermodynamic systems with
a characteristic Hawking temperature and an entropy, which, upto leading
order, is proportional to the area of the event horizon in Planck units
\cite{waldlex}. The entropy is a function of the mass (internal energy)
$M$, charge $Q$ and the angular momentum $J$ for the most general charged
rotating black holes. These serve as extensive thermodynamic variables
provided we consider the black hole to be a subsystem of a larger
thermodynamic system with which it is in equilibrium. 

A large class of 
extremal BPS black holes occur in the low energy supergravity theories
arising from string theory.  In particular the entropy of extremal black holes in
supergravity theories with $N\geq 2$ are known to posses an underlying
microscopic statistical description in terms of D-brane systems or
fundamental string states.  Although such extremal black holes have zero
Hawking temperature, they have a non zero thermodynamic entropy and are
described by degenerate quantum ground states. A microscopic state counting
in the associated conformal field theory then reproduces the thermodynamic
entropy as an asymptotic expansion in
the large charge limit. The low energy effective action of $N\geq 2$
supergravity following from type II string compactifications involve higher
derivative terms in an ${\alpha}'$ expansion. These terms
modify the Bekenstein Hawking area law and introduces subleading
corrections to the entropy. These corrections may be
computed from the Wald formulation of generally covariant higher derivative
theories of gravity \cite {waldlex}. It has been possible in the recent past to account for
these subleading corrections from a microscopic perspective following from
the underlying string theory. Exact matching between the macroscopic and
microscopic entropy upto various subleading orders in an asymptotic
expansion have been obtained for diverse extremal black holes
\cite{attractors}.

On the other hand, 
thermodynamic systems in equilibrium are known to possess interesting
extrinsic geometrical features \cite{callen} although intrinsic geometric
structures in equilibrium thermodynamics were largely unknown. An 
equilibrium state space of a thermodynamic system may be
considered to be described by the minima of the internal energy function $U=U\left(S/T, V/T,
\mu_i/T\right)$ in the energy representation or the 
maxima of the entropy $S=S\left(U, V, N_i\right)$ in the entropy 
representation.  Here the quantities $\mu_i, T, V,
N_i$ are the chemical potentials, temperature, volume and particle number
per species respectively. Weinhold \cite{wein} introduced an inner product
in this thermodynamic state space in the energy representation as the
Hessian matrix of the internal energy with respect to the extensive
variables leading to an intrinsic positive definite Riemannian geometric
structure.  One of the extensive parameters, typically the volume, was held
fixed to provide a physical scale and prevent the development of negative
eigenvectors of the metric. Although interesting, the physical relevance of this
structure seemed elusive. Ruppeiner \cite{rup1} reformulated the Weinhold
inner product in the entropy representation in terms of the 
(negative of the)
Hessian matrix of the entropy with respect to the extensive thermodynamic
variables. This also led to a positive definite Riemannian geometric
structure in the thermodynamic state space which was conformally related to
the Weinhold geometry, with the temperature as the conformal factor.
Ruppeiner showed that consideration of thermodynamic fluctuation theory
\cite{landau} in addition to the thermodynamic laws allowed a remarkable
physical interpretation of this geometric structure in terms of the probability 
distribution of the fluctuations and a relation of the scalar curvature with 
critical phenomena. 

Thermodynamic geometry of the equilibrium state space described above may
also be applied to study black holes considered as thermodynamic systems.
Recent studies of the thermodynamics of diverse black holes in this
geometric framework have elucidated interesting aspects of phase
transitions and relations to moduli spaces of $N\geq 2$ supergravity
compactifications in the context of extremal black hole solutions in these
theories. It may be argued, however, that the connection of this formulation
to fluctuation theory for application to black holes requires several
modifications \cite{rup3}.  The geometric formulation in the state space
was first applied to $N\geq 2$ supergravity extremal black holes in $D=4$
which arise as low energy effective field theories from compactification of
Type II string theories on Calabi-Yau manifolds \cite{fgk}. Since then,
several authors have attempted to understand this connection \cite{cai1},
\cite{aman123} both for supersymmetric as well as
non-supersymmetric extremal and non extremal black holes and five dimensional rotating black rings. 
In this context, we had explored the state space geometry of
both non extremal rotating BTZ black holes and BTZ-Chern Simons (BTZ-CS)
black holes in the Ruppeiner formulation \cite{sgt}. 

It is a natural question to pose whether the formalism of thermodynamic
geometry of the equilibrium state space can be generalised to systems at zero temperature. 
Black holes
at extremality provide the most natural laboratory to study this issue.
Clearly, notions of conventional thermodynamics are not expected to be valid
in this regime, and one would require modifications to the same. The
first question that one can address, however, is whether a geometric
characterization of the state space is at all possible at extremality. 
It is one of the issues that we will address in this
paper, and we will see that this is indeed the case. 

In this paper, we focus our attention on the charged extremal BPS black
holes in Type II string theories compactified on Calabi-Yau manifolds.
As is well known by now, the radial variation of the moduli in these
cases exhibit attractor behaviour as they flow from their asymptotic values
to an attractor fixed point at the horizon where these are fixed in terms of the charges
through the attractor equations. 
These BPS black hole solutions (of $N=2$ supergravity) fall in two distinct
classes : namely the large black holes which have a non vanishing area
at the two derivative level and posses dyonic charges, and the small
black holes which have a vanishing area and carry electric charges only in
a suitable duality basis. The large black holes may be described in
terms of wrapped branes on non-trivial cycles of the compact internal
manifold. Their microscopic entropy is determined in terms of the
microstate counting through the Cardy formula in the underlying two
dimensional CFT associated with the brane system. This
is in precise agreement upto the subleading terms with
the macroscopic entropy following from the Wald formula.  The small
black holes are more complicated. They have a vanishing horizon area as
the horizon coincides with the null singularity and the curvature diverges.
Hence higher curvature terms are large and the singularity is cloaked by
the effective horizon \cite{attractors}. Although a lot of progress has
been made in understanding these systems, there are still certain unresolved issues.

The macroscopic entropy also follows directly from a variational principle applied to a
generic class of {\it entropy functions} of the charges and the moduli. The attractor equations
follow from the extremisation of this function and lead to the attractor fixed point at the horizon.
An alternative analysis due to Sen \cite {senall} involves an adaptation of the Wald
formalism to establish a more general variational technique to compute the
higher derivative corrections to the entropy of charged extremal black
holes. This formalism involves a more general class of entropy functions
which are functions of both the scalar moduli and the parameters which
describe the near horizon $AdS_2 \times S^{(D-2)}$ geometry. Extremisation
of this entropy function determines all the near horizon parameters and it
maybe shown that the entropy function at the attractor fixed point
determines the black hole entropy. Higher derivative contributions to the
entropy may also be elegantly implemented through this general entropy
function formalism. Typically, the generalised
entropy function formalism is mostly independent of supersymmetry
considerations and has been also applied to extremal but non supersymmetric
black holes \cite{senextra}.

The entropy of extremal black holes which have zero  
Hawking temperatures naturally alludes to a non trivial limiting 
characterisation of conventional thermodynamics. This is further 
supported by the gauge-gravity correspondence in which certain limiting 
thermodynamic notions emerge for describing extremal black holes
\cite{shiraz1}. The entropy arises from macroscopic degeneracy of a 
quantum ground state and is a familiar phenomena in the physics of 
condensed matter like spin glasses. A thermodynamic interpretation of 
this macroscopic degeneracy may be formally attempted through the 
partition function in a grand canonical ensemble involving summation 
over the chemical potentials. Although an exact evaluation of this 
formal expression is difficult it maybe possible within the
gauge gravity correspondence to compute the sum in the boundary gauge 
theory \cite{shiraz2}. An alternative approach to a limiting zero 
temperature characterisation of thermodynamics also arises from the 
general AdS-CFT correspondence \cite{deboer}.

The possibility of such a limiting characterisation of thermodynamics at 
zero temperatures naturally leads to the question of the geometry of 
the equilibrium thermodynamic state space of extremal black holes. 
This is a reduced state space of 
equilibrium thermodynamic states consistent with the extremality 
condition. For supersymmetric black holes this is simply the BPS 
condition although the general arguments should also hold for non 
supersymmetric extremal black holes. As geometric notions remain valid even 
at extremality, a non degenerate geometric realization of the state space
of extremal black holes should be possible within the framework of 
thermodynamic geometries. The possibility of
a thermodynamic geometry for extremal black holes was first alluded to
in the context of  Type II supergravities with special geometry of the 
moduli space \cite{fgk}. In accordance with the first law of thermodynamics, 
the equilibrium state space for 
these black holes would now also involve the moduli fields 
at asymptotic infinity apart from the electric and 
the magnetic charges \cite{kol} as extensive thermodynamic variables. 
This, in general, leads to a curved  equilibrium thermodynamic state space. 

Although conventional thermodynamics is not valid at extremality, the
geometric features of the state space should continue to be valid and well
defined. However, the connection to thermodynamic fluctuations is elusive
even for non extremal black holes \cite {rup3} and is certainly not expected 
to hold in the zero temperature extremal limit. In fact its well known that 
classical fluctuations which have a thermal origin ceases at zero temperatures.
The scalar curvature on the other hand may possibly still indicate
interactions, and divergences in the scalar curvature may allude to zero
temperature quantum phase transitions amongst distinct vacua in the moduli
space. In particular a generalized thermodynamic
geometry of the equilibrium state space extended by the moduli variables
should lead to insights into thermodynamics away from the attractor fixed
point. This is also expected to provide a geometrical understanding of the
attractor mechanism from flows in the space of thermodynamic geometries and
a geometric comprehension of the attractor fixed point. As a first
step toward this objective, it is necessary to explore the state space
geometry for extremal black holes at the attractor fixed point. In
particular, it is important to explore whether thermodynamic metrics are non
degenerate at extremality and to establish the behavior of the scalar
curvature. Such a construction would clearly elucidate the issue of
thermodynamics at extremality and provide a geometrical realization of the
equilibrium thermodynamic state space at zero temperatures. It would also be
interesting to study the effect of the higher derivative contributions to
the extremal black hole entropy, on the scalar curvature in the state space.
It is possible that the higher derivative corrections may induce a
modification of the divergences in the scalar curvature in the state space
and consequently that of possible quantum phase structures.
Small black holes in Type II supergravity which have zero entropy at the two derivative level
are particularly interesting in this
connection as the thermodynamic geometry arises only from the higher
derivative contributions to the entropy.

We emphasize here however that since we are considering a limiting characterization
of thermodynamics arising from the macroscopic degeneracy of a quantum ground state,
conventional thermodynamic notions will need to be modified. In particular the usual
relation of the geometry of equilibrium state space with thermodynamic fluctuations
is not expected to be valid at zero temperatures. 
The characterization of the probability distribution of 
thermodynamic fluctuations in terms of an invariant positive definite Riemannian form
over the equilibrium state space is not expected to hold for extremal black holes.
So although we expect the emergence of a non degenerate geometric structure of the state
space through our construction, positive definiteness of the Riemannian form is not 
a strict criterion. However, the usual notion of thermodynamic stability in the canonical ensemble
requiring the positivity of specific heats or compressibilities are still expected to hold.
In fact, the invariant Riemannian form on the thermodynamic state space for such systems can
be indefinite and may even be sensitive to the higher derivative corrections to the entropy.
Although it is premature to make a definite statement, the attractor mechanism for extremal black holes seems to suggest a role for the moduli space metric in determining the signature of the Riemannian form.

In this paper, we examine the geometry of the equilibrium thermodynamic
state spaces of three different charged extremal black holes in Type II
supergravity with 4 charges, 3 charges and small black holes with two
charges. These are described respectively by microscopic D2-D6-NS5-P
and D1-D5-P brane systems and Type IIB string theory compactified on
$K_3\times T^2$. The paper is organised as follows. 
In the next section we provide a brief review of
thermodynamic geometries of the equilibrium state spaces of thermodynamic
systems and its relation to interactions and phase transitions both for two
and higher dimensional thermodynamic state spaces. In section 3, we
explore the thermodynamic geometry of the four charged extremal black hole
arising from a microscopic configuration of D2-D6-NS5-P brane system
and further study the modification to the thermodynamic geometry from
higher derivative contributions to the entropy. In section 4, we investigate
the state space geometry of the three charged extremal black hole arising
from a D1-D5-P system both in $D=5$ and $D=10$ and also illustrate the
effect of higher derivative contributions to the entropy on the state space
geometry and its curvature. In section 5, we take up the
interesting issue of two charged small black holes in Type IIB string
theory compactified on $K_3 \times T^2$ and explore the thermodynamic
geometry of the state space arising out of both the macroscopic and the
microscopic entropy expression resulting from a heterotic string theory
computation. We further consider the modification of the state space
geometry due to the higher derivative corrections to the entropy both from
the low energy effective action as well as string loop corrections. We also
study the state space geometry and its curvature implied by an exact
entropy expression following from a quantum information theory perspective.
Section 6 concludes this paper with a summary of the results.

\section{Review of Thermodynamic Geometry}

In this section we present a brief review of the essential features of
thermodynamic geometries and their application to the thermodynamics of
black holes, in particular extremal black holes. This will serve to to 
set the notations and conventions used in the rest of this paper. 
An intrinsically geometric structure in equilibrium thermodynamics 
was introduced by Weinhold \cite{wein} through an inner product in the space of
equilibrium thermodynamic macrostates defined by the minima of the internal
energy function $U=U(S/T, V/T, \mu_i/T)$ as the Hessian
\bea
h_{ij}=\pt_i\pt_j U
\eea
As mentioned in the introduction, 
the quantities $\mu_i, T, V, S$ are the chemical potentials,
temperature, volume and entropy respectively and the volume or any other
parameter is held fixed to provide a physical scale and to restrict
negative eigenvectors of the metric. Although such a Riemannian geometric
structure was interesting, no physical significance could be ascribed to
it. The inner product on the state space was later reformulated by
Ruppeiner \cite{rup1} in the entropy representation as the negative of the
Hessian matrix of the entropy with respect to the extensive variables. The
thermodynamic macrostates underlying the equilibrium state space being now
described by the maxima of the entropy function $S =S(U, V, N)$. Explicitly
the Ruppeiner metric in the state space was given as
\bea
g_{ij}=-\pt_i\pt_j S( U, V, N)
\eea 
and was conformal to the Weinhold metric with the inverse temperature as
the conformal factor.  The negative sign was necessary to ensure positive
definiteness of the metric, as the entropy is a maximum in the equilibrium
state.  It could be shown that the Riemannian structure defined by the
Ruppeiner metric was closely related to classical thermodynamic fluctuation
theory \cite{rup2} and critical phenomena.  The probability
distribution of thermodynamic fluctuations in the equilibrium state space
was characterised by the invariant interval of the corresponding
thermodynamic geometry in the Gaussian approximation as
\begin{equation}
W(x) = A {\rm exp}\left[ -\frac {1}{2} g_{ij}(x) dx^i dx^j\right]
\end{equation} 
where $A$ is a constant. The inverse metric may be shown to be the second
moment of fluctuations or the pair correlation functions and given as
$g^{ij}= < X^i X^j>$ where $X^i$ are the intensive thermodynamic variables
conjugate to $x^i$.  The Riemannian structure may be expressed in terms of
any suitable thermodynamic potential arrived at by Legendre transforms
which corresponds to general coordinate transformations of the equilibrium
state space metric. The geometric formulation tacitly involves a statistical
basis in terms of a canonical ensemble although the analysis is considered
only in the thermodynamic limit.

For a standard two dimensional thermodynamic state space defined by the
extensive variables $ (x^1, x^2)$, application of these geometric
notions to conventional thermodynamic systems suggest that a non zero scalar
curvature indicates an underlying interacting statistical system. It may be
shown that the scalar curvature $R\sim\kappa_2\xi^d$ where $\xi$ is the
correlation length, $d$ is the physical dimensionality of the system and
$\kappa_2$ is a dimensionless constant of order one.   
The Ruppeiner formalism has been applied to diverse condensed
matter systems with two dimensional state spaces and is completely
consistent with the scaling and hyperscaling relations involving critical
phenomena and have reproduced the corresponding critical indices.

Having provided thus a brief account of thermodynamic geometries, in the following
sections we systematically explore the state space geometry in the Ruppeiner
framework for several charged extremal black holes in Type II string
theories. Our constructions
would provide a geometric realization of a limiting equilibrium
thermodynamics at zero temperatures. This would be the first step to address
the issue of the thermodynamics of extremal black holes away from the attractor fixed point
and a geometric characterization of the attractor mechanism. 
It may also be a prelude to the application of the formalism of thermodynamic 
geometries to the study of zero temperature quantum phase
transitions amongst distinct vacua in the moduli space of string theory compactifications.
We should mention here that the role of the scalar curvature over the state space 
in the Ruppeiner formalism is not clear at this stage in the scenario being envisaged. However,
application of this geometric formalism to non extremal black holes and the consequent 
divergences of the same at known critical points \cite{aman123} indicates that a non zero 
scalar curvature for state spaces of extremal black holes may also suggest an underlying 
interacting statistical system. In this case, the divergences of the scalar curvature may 
allude to zero temperature quantum phase transitions amongst distinct vacua in the 
moduli space. The attractor mechanism for extremal black holes and the consequent flow of 
the scalar moduli to fixed values at the horizon in terms of the charges also seems to suggest 
such a connection of the state space scalar curvature with the structure of the moduli space. 
It is this perspective that we will adopt in the present study and despite all the 
caveats, we will explore the scalar curvatures over the state space of extremal black holes 
and their sensitivity to higher derivative corrections to the entropy. Following the 
arguments presented earlier we will interpret the scalar curvature as indicative of an 
interacting microscopic statistical system underlying extremal black holes at zero 
temperatures.

\section{Four Charged Black Holes in D2-D6-NS5-P System.}

In this section, as a first exercise, we study the thermodynamic geometry of
four charged extremal black holes in $D=4$ and $D=10$ in Type II A string
theory compactified on $T^6$ \cite{malda}. These are known to be described
by microscopic D2-D6-NS5-P brane systems which are $\frac {1}{8}$ BPS
configuration in Type IIA supergravity.  The solitonic NS5 brane is
required for satisfying the tree level IIA supergravity equations of motion
and this does not affect the overall supersymmetry. From a microscopic
perspective this is analogous to exciting left moving oscillations on a
fundamental heterotic string.  The brane system in question is T dual to
the D1-D5-P system considered in the next section. The D2 branes are
sources for electric fields whilst the D6 and NS5 branes are sources
of magnetic fields. For the relevant expressions for the metrics in the 
various cases, see, e.g. \cite{d2d6}.  

\subsection{Four Charged Black Holes in $D=4$}

The macroscopic entropy at the tree level $\alpha\prime$
resulting from the two derivative part of the action is
\bea
S(N_2, N_5, N_6, N_p)= 2 \pi \sqrt{N_2 N_5 N_6 N_p}
\eea 
where, in the D-brane description, 
$N_2$, $N_6$ and $N_5$ are identified with the number of 
D-2, D-6 and NS-5 branes respectively, and $N_p$ with the number of 
units of Kaluza Klein momenta, and are assumed to be large.  
This is the standard Bekenstein-Hawking entropy given by the area law.
It is now possible to explore the thermodynamic geometry of the equilibrium
state space of the 4 charged extremal black hole in $D=4$ arising from this
entropy expression. The Ruppeiner
metric in the state space is given by the Hessian matrix of the entropy
with respect to the extensive variables which in this case are the four
conserved charges carried by the extremal black hole. A straightforward
computation yields

\begin{eqnarray}
 ds^2&=& \frac{\pi}{2}\left( 
\sqrt{ \frac{N_5 N_6 N_p}{N_2} } \frac{dN_2^2}{N_2} 
+  \sqrt{ \frac{N_2 N_6 N_p}{N_5} } \frac{dN_5^2}{N_5}  
+  \sqrt{ \frac{N_2 N_5 N_p}{N_6} } \frac{dN_6^2}{N_6}\right.\nonumber \\ 
&+&\left. \sqrt{ \frac{N_2 N_5 N_6}{N_p} } \frac{dN_p^2}{N_p}\right)
-  \pi \left(\sqrt{\frac{N_6 N_p}{N_2 N_5}}dN_2 dN_5 
+  \sqrt{\frac{N_5 N_P}{N_2N_6}}dN_2 dN_6 \right.\nonumber \\
&+& \left.\sqrt{\frac{N_5N_6}{N_2N_p}}dN_2 dN_p 
+  \sqrt{\frac{N_2 N_p}{N_5 N_6}}dN_5 dN_6 
+  \sqrt{\frac{N_2 N_6}{N_5 N_p}}dN_5 dN_p \right.\nonumber \\
&+& \left.\sqrt{\frac{N_2 N_5}{N_6 N_p}}dN_6 dN_p \right) 
\end{eqnarray}
The determinant of the metric tensor is $ g= -\pi^4 $ which is a constant
and hence we have a non degenerate metric for the thermodynamic state space
at extremality. Note that the determinant is of negative sign, implying that
the metric is {\it not} positive definite. As we have emphasized that in the case of
extremal black holes being considered, positivity of the thermodynamic metric is
not a strict criteria as classical fluctuations of thermal origin are absent at zero temperatures.

The fact that the metric is non-degenerate is somewhat remarkable,
as conventional thermodynamics
breaks down at extremality. However our results seemingly indicates that
possibly a limiting characterisation at zero temperatures, of conventional
thermodynamics still holds at extremality. This possible
limiting thermodynamic description may arise from the fact that extremal
black holes are quantum ground states exhibiting macroscopic degeneracy.
The non degenerate metric at extremality thus provides a geometric
realization of the thermodynamic state space of the extremal black hole at
zero temperature.

Note that the state space geometry is four dimensional as the thermodynamic
entropy is a function of four charges which serve as extensive variables.  
As mentioned earlier, classical fluctuations are absent
at zero temperatures although the curvature scalar possibly still continues
to indicate an underlying interacting statistical system given the fact that
it does so for non extremal black holes.  The curvature scalar over the state space 
in this case may be easily computed to be
\bea
R = \frac{3}{2 \pi \sqrt{ N_2 N_5 N_6 N_p }} 
\eea
Thus, the curvature scalar is non zero, finite and regular everywhere although in
the large charge limit at which the entropy computation is valid the
curvature is small. This should indicate an underlying weakly interacting
statistical basis. The entropy considered here is the one
arising from the usual two derivative terms in the low energy effective
supergravity action which is consistent with the area law. This 
is modified by contributions from the higher
derivative terms in the effective action, and consequently modifies
the equilibrium state space and the thermodynamic geometry including the
curvature scalar. In what follows we will consider such modifications to
the entropy of the $D=10$ four charged black holes and study the consequent
thermodynamic geometries of the equilibrium state space based on the
modified entropy expression.

\subsection{ Four Charged Black Holes in $D=10$}

The near horizon geometry of
the $D=10$ extremal charged black holes described by the D2-D6-NS5-P
system is $AdS_3 \times S^2\times S^1\times T^4$. 
The Wald formula in the framework of the Sen entropy function with some
modifications for the non standard near horizon geometry leads to the
entropy of the four charged extremal black hole in $D=10$ at the two
derivative level as 
\begin{equation} 
S(N_2, N_5, N_6, N_p)= 2 \pi \sqrt{N_2 N_5 N_6 N_p}
\end{equation}
with the $N_i$s defined as before. 
This is identical to the $D=4$ case and follows the standard area law
and the universal expression for entropy of charged extremal black holes as
the square root of the product of the charges. The thermodynamic geometry
of the equilibrium state space arising from the entropy at the two
derivative level is hence also identical to the $D=4$ case.

\begin{figure}
\centering
\epsfxsize=3.5in
\hspace*{0in}\vspace*{0.2in}
\epsffile{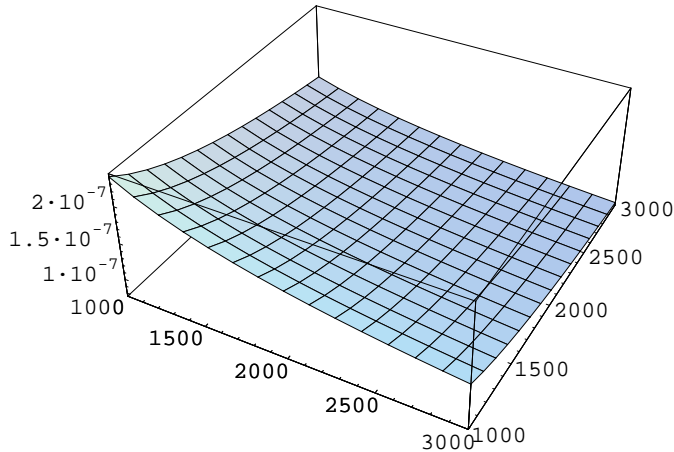}
\caption{\small Ruppeiner curvature for the D2-D6-NS5-P System as a 
function of $N_2$ (X-axis) and $N_5$ (Y-axis).}
\label{d2d6ns5p}
\end{figure}

It is possible to consider the subleading corrections to the entropy of the
four charged extremal black hole in $D=10$ following from the contribution
of the higher derivative terms in the low energy effective action. The Sen
entropy function framework is applicable for this exercise in spite of the
non standard horizon geometry, with certain modifications. 
The corrected entropy actually depends on a parameter which
involves field redefinitions for the higher derivative terms \cite{d2d6}.
This essentially happens for the D2-D6-NS5-P system because the
supergravity configuration admits a near horizon geometry involving $AdS_3$
and $S^2$ with different radii (unlike the D1-D5-P system considered
later). The ambiguity due to field redefinitions may be used to define a
specific redefinition scheme in which only the Weyl tensor part of the
curvature occurs in the higher derivative terms. The $\alpha \prime$
corrected entropy for the four charged extremal black hole may now be
computed using the Wald formula and the Sen entropy function extremisation
framework. An explicit computation gives the corrected entropy as \cite{d2d6}
\begin{equation}
S(N_2, N_5, N_6, N_p)= 2 \pi \sqrt{N_2 N_5 N_6 N_p} - 2\pi C
\frac{\sqrt{N_p}}{N_2 N_6 N_5^{5/2}} 
\end{equation}
where $C = \frac{73315}{222184} \left(\frac{2R_4}{G_N^4 \alpha\prime R_9}\right)^{3/2}\gamma$
where $R_4$ and $R_9$ refer to the radii of the circles over which the D-2 brane is
wrapped, $G^4_N$ is the 4-D Newton's constant, and $\gamma = \frac{1}{8}\zeta(3)\alpha'^3$ 
The thermodynamic geometry of the equilibrium state space for the four
charged extremal black hole resulting from the corrected entropy may now be
computed from the Hessian matrix of the entropy with respect to
the charges as :

\begin{eqnarray}
 ds^2&=&  
\left(\frac{\pi}{2N_2}\sqrt{\frac{N_5 N_6 N_p}{N_2}}+ 
   \frac{4\pi C \sqrt{N_p}}{N_2^3N_6N_5^{5/2}}\right)dN_2^2
-\left(\pi \sqrt{\frac{N_6 N_p}{N_2 N_5}}- 
   \frac{10\pi C \sqrt{N_p}}{N_2^2N_6N_5^{7/2}}\right)dN_2 dN_5\nonumber\\
&-& 
\left(\pi \sqrt{\frac{N_5 N_P}{N_2N_6}}- 
    \frac{4\pi C\sqrt{N_p}}{N_2^2N_6^2N_5^{5/2}}\right)dN_2 dN_6
-\left(\pi \sqrt{\frac{N_5N_6}{N_2N_p}}+
   \frac{\pi C}{\sqrt{N_p}N_2^2N_6N_5^{5/2}}\right)dN_2 dN_p\nonumber\\
&+&\left(\frac{\pi}{2N_5} \sqrt{ \frac{N_2 N_6 N_p}{N_5} }+
   \frac{35\pi C\sqrt{N_p}}{2N_2N_6N_5^{9/2}}\right)dN_5^2
- \left(\pi \sqrt{\frac{N_2 N_p}{N_5 N_6}}-
    \frac{10\pi C \sqrt{N_p}}{N_2^2N_6N_5^{7/2}}\right)dN_5 dN_6 \nonumber\\
&-&\left(\pi \sqrt{\frac{N_2 N_6}{N_5 N_p}}+
 \frac{5\pi C}{\sqrt{N_p}N_2N_6N_5^{7/2}}\right)dN_5 dN_p+
\left(\frac{\pi}{2N_6}\sqrt{ \frac{N_2 N_5 N_p}{N_6}}+
  \frac{4\pi C\sqrt{N_p}}{N_2N_6^3N_5^{5/2}}\right)dN_6^2\nonumber\\
&-& \left(\pi \sqrt{\frac{N_2 N_5}{N_6 N_p}}+
 \frac{2\pi C}{\sqrt{N_p}N_2N_6^2N_5^{5/2}}\right)dN_6 dN_p\nonumber\\
&+&\left(\frac{\pi}{2N_p}\sqrt{ \frac{N_2 N_5 N_6}{N_p} }
-\frac{\pi C}{2N_p^{3/2}N_2N_6N_5^{5/2}}\right)dN_p^2 
\end{eqnarray}
The determinant of the metric tensor is 
\begin{eqnarray}
g &=& \frac{-\pi^4}{N_2^6\sqrt{N_p}N_5^{12}N_6^6} 
(6C^2\sqrt{N_p}N_2^3N_5^6N_6^3+ 
5CN_2^4N_5^{17/2}N_6^4 \sqrt{N_2 N_5 N_6 N_p}\nonumber\\
&+& 50C N_2N_5^{5/2}N_6\sqrt{N_2 N_5 N_6 N_p}+ 
N_2^6\sqrt{N_p}N_5^{12}N_6^6+ 100C^4\sqrt{N_p})
\end{eqnarray}
and is non zero in the large charge limit indicating a non degenerate
thermodynamic geometry at extremality modified by the higher derivative
contributions. The scalar curvature of the thermodynamic state space may
now be computed in a straightforward fashion. The explicit expression for
the scalar curvature is long and is hence relegated to the
Appendix. The scalar curvature is regular in the large
charge limit in which the asymptotic expansion for the entropy is valid.
We provide a graphical analysis of the scalar curvature for different
values of the charges in fig. (\ref{d2d6ns5p}). In the figure, we have shown the 
Ruppeiner scalar curvature of the D2-D6-NS5-P system, as a function of 
$N_2$ (X axis) and $N_5$ (Y axis), both of which vary from 
$10^3$ to $3 \times 10^3$. To illustrate the example, 
we have chosen the typical values $N_6 \simeq N_p = 2\times 10^3$. 
The value of the constant $C$ has been set to $10^{-5}$. We notice
a small but positive Ruppeiner curvature at these typical values of
$N_2$ and $N_5$.  

\section{Black Holes in the  D1-D5-P System}

In this section we explore the thermodynamic geometry of 3 charged extremal
black holes in Type IIB string theory arising from a microscopic
configuration of $N_1$ D1 branes wrapped along some compact direction
$y$ with radius $R$, $N_5$ D5 branes wrapped along $y$ and a four
torus $T^4$, and
$p=N_p/R$ units of KK momentum along the $y$ direction. The corresponding
five dimensional extremal black holes carry both electric and magnetic
charges arising from the D1 and D5 branes respectively. In $D=10$ the
near horizon geometry of the configuration is $M_3 \times S^3 \times T^4$
where $M_3$ is a boosted $AdS_3$ geometry. The ten dimensional metric may
be reduced over $S^1\times T^4$ and the resulting $D=5$ charged extremal
black hole involves a more tractable near horizon geometry of $AdS_2\times
S^3$. The near horizon geometries imply that the standard entropy function
method may be applied to compute the entropy of these extremal charged
black holes at the two derivative level. Higher derivative contributions
are however tricky as the non standard near horizon geometry for the
corresponding configuration in $D=10$ complicates the application of the
Wald formula. Despite this the $\alpha^{3 \prime} R^4$ corrections arising
from string tree level scattering amplitudes have been incorporated to
compute the subleading corrections to the entropy of extremal charged black
holes in such D1-D5-P systems in $D=10$ \cite{d1d5}.

We now explore the thermodynamic geometry of these D1-D5-P
extremal charged black holes in $D=5$ and $D=10$.  We further elucidate the
modification of the thermodynamic geometry involving the higher order
$\alpha^{3\prime}R^4$ corrections in the $D=10$ type IIB supergravity effective
action, and their subleading contributions to the entropy

\subsection{D1-D5-P black holes in $D=5$.}

The $D=10$ IIB supergravity may be compactified to $D=5$ on $S^1 \times
T^4$.  The near horizon limit of the metric is $AdS_2\times S^3$ allowing 
the application of the entropy function formalism
to compute the Bekenstein-Hawking entropy as 
\be
S_{BH} = 2 \pi \sqrt{N_1 N_5 N_p}.
\ee
The result coincides with that obtained by direct computation of the
horizon area verifying the usual area law at the two derivative level.

The metric of the thermodynamic state space in the entropy representation
may now be obtained as before from the Hessian matrix of the entropy with
respect to all the extensive thermodynamic variables which in this case are
just the D1, D5 and P charges. Explicitly the metric is given as

\begin{eqnarray}
dS_{BH}^2&=& \frac{\pi}{2}\left(
\sqrt{ \frac{N_5 N_p}{N_1} } \frac{dN_1^2}{N_1} + 
\sqrt{ \frac{N_1 N_p}{N_5} } \frac{dN_5^2}{N_5} \right.\nonumber\\
&+&\left.\sqrt{ \frac{N_1 N_5}{N_p} } \frac{dN_p^2}{N_p}\right) 
- \pi\left(\sqrt{ \frac{N_p}{N_1 N_5} } dN_1 dN_5 \right.\nonumber\\
&+& \left.\sqrt{ \frac{N_5}{N_1 N_p} } dN_1 dN_p +
        \sqrt{ \frac{N_1}{N_5 N_p} } dN_5 dN_p\right) 
\end{eqnarray}

We observe that as in the previous example the Ruppeiner metric is non
degenerate and regular everywhere even at extremality. The determinant of
the metric tensor is 
\begin{equation}
g= -\frac{\pi^3}{2\sqrt{N_1 N_5 N_p}} 
\end{equation}
and is non zero for non zero charges. \footnote{As in the previous 
subsection, the metric is not  positive definite, but our previous 
argument will still hold in this case, and this is not an issue.}  
This lends credence to the contention of a
limiting characterisation of conventional thermodynamics at zero
temperature, in this case. The Ruppeiner metric at extremality obtained by us thus
characterises a geometric realization of the corresponding equilibrium
thermodynamic state space at extremality. 

The scalar curvature corresponding to the geometry of the thermodynamic 
state space may now be determined to be
\be
R = \frac{3}{4 \pi \sqrt{ N_1 N_5 N_p }} 
\ee

We observe that the scalar curvature is non zero, positive and regular
everywhere. As mentioned in connection with the previous example of the
four charged extremal black holes in Type II supergravity, this fact seems
to be universal and is related to the typical form for the Ruppeiner
geometry as the Hessian matrix of the entropy. The
usual connection of Ruppeiner geometry with thermodynamic fluctuation
theory needs to be modified in connection with the application to black
holes \cite {rup3}. However the standard interpretation of scalar curvature of the state
space geometry to describe interactions in the underlying statistical
system should continue to hold for non extremal black holes. In particular this should
also hold at extremality and hence the non zero scalar curvature for the
state space of D1-D5-P system should indicate an underlying interacting
statistical basis. 

\subsection{ D1-D5-P  black holes in $ D=10$.}

The black hole solution corresponding to a microscopic D1-D5-P brane system
maybe described by a modified non standard near horizon $AdS_3 \times S^3
\times T^4$ geometry. For this case, the horizon at
$r=0$ has a geometry of $S^1\times S^3 \times T^4$. This complicates the
application of the Sen entropy function approach to compute the black hole
entropy. The relations between the number of D-branes with constants of
integration and the ADM momentum in $y$ needs to be computed again due to
the modified near horizon geometry. Furthermore the Wald formula for this
modified background needs to be recomputed and is involved due to the
presence of an off diagonal term in the metric describing the modified near
horizon geometry. Fortunately the computation for the Wald formula may be
simplified by using certain properties for the Riemann tensor of the
modified near horizon geometry. This enables the application of the Sen
entropy function method and the entropy may be computed to leading order or
two derivative level to be \cite{d1d5} 
\begin{equation} 
S(N_1, N_5, N_P)= 2\pi \sqrt {N_1 N_5 N_P}
\end{equation}
which is 
exactly the same as in the $D=5$ case. Hence the Ruppeiner geometry of the
thermodynamic state space for the $D=10$ D1-D5-P black hole is
identical to the $D=5$ case. The scalar curvature of the state space is
also identical and exactly the same interpretations should continue to hold
as in the $D=5$ case.

As in the instance of the D2-D6-NS5 brane system, the entropy
expression for the system in question would also be modified by the
subleading contributions from the higher derivative terms in the low energy
Type IIB supergravity action. Following exactly the same approach, at the
next leading order in $\alpha'$ (with spacetime $R^4$ corrections)
arising out of the string three loop scattering amplitudes, the entropy of
the extremal black hole is given as 
\begin{equation} 
S_{BH} = 2 \pi \sqrt{N_1 N_5 N_p}\left[1 + \frac{b}
{\left(N_1 N_5\right)^{3/2}}\right]
\label{corr1}
\end{equation} 
where $b=3\gamma\left[\frac{2 \pi^3 V_4}{16\pi G_{10}}\right]^{3/2}$ 
with $G_{10}$ being the $10$ dimensional Newton's coupling constant, 
$V_4$ the volume of the $T^4$ and $\gamma=\frac{1}{8}\zeta
(3)\alpha^{\prime 3}$.

The Ruppeiner metric may now be computed as the Hessian matrix of the
corrected entropy expression in eq. (\ref{corr1}) to be,
\begin{eqnarray}
 ds^2 &=& 
\left[\frac{\pi}{2N_1}\sqrt{\frac{N_5 N_p}{N_1}}\left(1+ 
\frac{b}{(N_1 N_5)^{3/2}}\right)- 
\frac{9\pi b \sqrt{N_p}}{2 N_1^3 N_5}\right] dN_1^2\nonumber\\
&-&\left[\frac{\pi N_p}{\sqrt{N_1 N_5 N_p}}\left(1+ 
\frac{b}{(N_1 N_5)^{3/2}}\right)+ 
 \frac{3\pi b\sqrt{N_p}}{(N_1 N_5)^2}\right] dN_1 dN_5 \nonumber\\
&-&\left[\frac{\pi N_5}{\sqrt{N_1 N_5 N_p}}\left(1+
\frac{b}{(N_1 N_5)^{3/2}}\right)-\frac{3\pi b}{\sqrt{N_p}N_1^2 N_5}\right] 
dN_1 dN_p\nonumber\\
&+&\left[\frac{\pi}{2N_5}\sqrt{\frac{N_1 N_p}{N_5}}\left(1+ 
\frac{b}{(N_1 N_5)^{3/2}}\right)-\frac{9\pi b\sqrt{N_p}}{2N_1N_5^3}\right] 
dN_5^2\nonumber\\
&-&\left[\frac{\pi N_1}{\sqrt{N_1 N_5 N_p}}\left(1+ \frac{b}{(N_1 N_5)^{3/2}}
\right)-\frac{3\pi b}{\sqrt{N_p}N_1 N_5^2}\right]dN_5 dN_p\nonumber\\
&+&\frac{\pi}{2N_p}\sqrt{\frac{N_1 N_5}{N_p}} 
\left(1+ \frac{b}{(N_1 N_5)^{3/2}}\right) dN_p^2 
\end{eqnarray}
The determinant of the metric tensor is given as 
\begin{equation} 
g= -\frac{\pi^3}{2N_1^5 N_5^5 \sqrt{N _p}}\left[\left(N_1 N_5\right)^{9/2}+ 
6b\left(N_1 N_5\right)^{3/2}- 20b^3\right] 
\end{equation}
The determinant is non zero for non zero charges and as before we obtain 
a non degenerate thermodynamic geometry of the state space of the three 
charged extremal black hole at extremality.

\begin{figure}
\centering
\epsfxsize=3.5in
\hspace*{0in}\vspace*{0.2in}
\epsffile{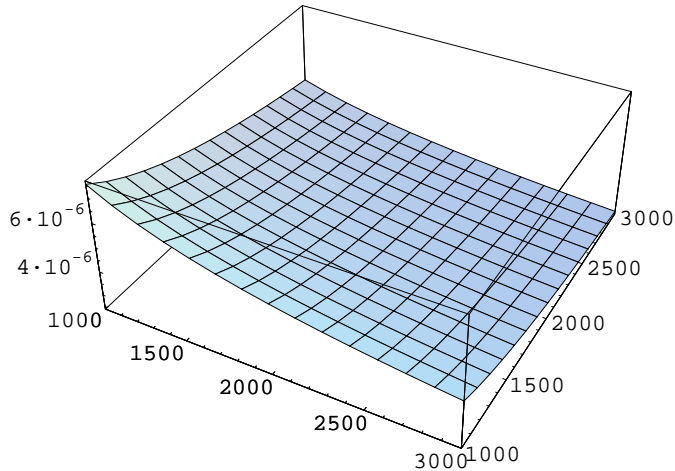}
\caption{\small Ruppeiner curvature for the D1-D5-P system with higher
derivative corrections, as a function of $N_1$ and $N_5$.}
\label{d1d5}
\end{figure}

The scalar curvature of the state space may now be computed using the
Ruppeiner metric based on the Hessian matrix of the entropy corrected by
higher derivative contributions. The curvature is non zero and regular
everywhere indicating again a stable thermodynamic system and underlying
interacting statistical basis. The exact expression for the scalar curvature
is involved and we relegate it to the Appendix. We trace the behavior of
the scalar curvature graphically in fig. (\ref{d1d5}), where we have 
plotted the Ruppeiner curvature as a function of $N_1$ and $N_5$, We have
set the typical values $N_p = 10^3$ and $b = 10^{-5}$. Of course, 
the scalar curvature is identical to that of the leading order case 
for $b=0$ for which the higher derivative contributions are absent.

\section{Two Charged Small Black Holes.}

In this section we study the third example of Ruppeiner geometry of the
thermodynamic state space of extremal charged {\it small} black holes in
string theory. These black holes
are characterised by a vanishing horizon area and hence zero entropy at the
two derivative level. However, higher derivative contributions renders this
entropy to be non zero at the next to leading orders and also provides a
finite non zero horizon area. Hence the naked singularity at the two
derivative level is cloaked by a horizon arising from the higher derivative
terms in the low energy effective supergravity action\cite{attractors}.
This renders the macroscopic description of small black holes somewhat
complicated as the higher derivative terms are now crucial for
the formation of a horizon. The microscopic statistical description of
small black holes are however simpler as it is based on just the
fundamental string states. 

\subsection {Macroscopic and Microscopic Description of Small Black Holes}

In Type IIA string theory compactified on $K_3 \times T^2$, small black
holes are charged objects in a suitably chosen duality basis.  
The macroscopic entropy in the two charged example of small black holes is 
\begin{equation}
S_{macro} = 4\pi\sqrt{\vert q_0 p^1 \vert}
\end{equation}
where $q_0$ and $p^1$ are the electric and magnetic charges respectively.
In the dual description, this small black hole is a charged black hole 
with charges $(q_0,p^1)$ of $N= 4$ heterotic string theory compactified 
on the six torus. It turns out that in this
case the radius of the horizon is of the string length $l_s$ and the small
black hole is at the {\it correspondence point} where it may be described
by a perturbative heterotic string state with the appropriate charge
configuration \footnote {We will use the notation $q_0=q$ and $p^1=p$ in what follows}. It is possible to
compute the degeneracy of these string states as the black hole is a $\frac
{1}{2}$ BPS configuration and hence corresponds to a short multiplet of the
$N=4$ supersymmetry. The degeneracy may be computed by using the Rademacher
formula and in the limit of large charges yields the microscopic entropy as
\bea
 S_{micro} &=& \ln{d(q,p)} 
\simeq  4\pi \sqrt{\vert qp \vert} - 27/4 \ln{\vert qp \vert} 
+O(\frac{1}{\sqrt{ \vert qp \vert}})\\
&\sim&  4\pi (s + \overline{s})_{Hor.} - 27/2 \ln{(s + \overline{s})_{Hor}}
\eea
where $s$ is the heterotic axion-dilaton field.
 
The macroscopic entropy matches the microscopic one at
leading orders but fails at subleading orders where the microscopic entropy
expression involves logarithmic corrections. So it follows that the
macroscopic entropy also must incorporate such corrections. 
It turns out that non holomorphic corrections are required to be
incorporated to $S_{macro}$ as the microscopic entropy expression involves
both $s$ and its complex conjugate $\overline {s}$.

The toroidally compactified heterotic string theory posses a duality
invariant spectrum but the entropy expression involving the generalized
holomorphic prepotential of the $N=2$ Lagrangian applied to toroidally
compactified Type II string theory is not duality invariant. The duality
invariance requires incorporation of non holomorphic terms to the
generalized prepotential. After incorporating the same,  
in both the attractor equations and the macroscopic
entropy, an S duality invariant expression is obtained which also
incorporates a logarithmic correction term $ -12 \ln(s +
\overline{s})_{Hor}$. Thus, the mismatch between
$S_{macro}$ and $S_{micro}$ persists even with the incorporation of the non
holomorphic corrections. It could be shown later \cite{senextra} that a
computation in the heterotic description involving a grand canonical
ensemble shows an exact match between the microscopic and the macroscopic
entropy, although a Type II description still seems to be problematic.

\subsection{Thermodynamic Geometry of Small Black Holes.}

The thermodynamic geometry of the equilibrium state space of two charged
small black holes in Type IIA string theory may now be constructed based on
the expressions for the microscopic and the macroscopic entropy. 
The horizon area for small black holes vanishes at the
two derivative level and hence the leading order contribution to the
macroscopic entropy is zero. The first contribution to the macroscopic
entropy hence arises at the subleading order from the higher derivative
contributions as 
\begin{equation}
S_1 = 4\pi \sqrt{qp}
\end{equation}
where $q$ and $p$ refer to the non-zero electric and magnetic charge of
the black hole. Now, the Ruppeiner metric
may be computed from the Hessian matrix of the entropy with respect to the
charges. The equilibrium thermodynamic state space in this case is two
dimensional as we have chosen a two charge configuration. Explicitly the
metric is given as
\begin{equation}
ds^2= \frac{\pi}{q} \sqrt{\frac{p}{q}}dq^2
- \frac{\pi}{\sqrt{qp}}dqdp
+ \frac{\pi}{p} \sqrt{\frac{q}{p}}dp^2 
\end{equation}
It can be seen that the Ruppeiner metric based on the
macroscopic entropy at the first subleading order, is degenerate and the
determinant of the metric $g=0$. Hence there is no viable thermodynamic
geometry at this order for the two charged extremal small black hole in
question.

Surprisingly the situation changes when we consider the duality invariant
macroscopic entropy corrected by the non holomorphic contributions from the
generalized prepotential. The macroscopic entropy
in this case is given as 
\begin{equation}
S(q,p)= 4 \pi \sqrt{qp}- 6\ln(qp) 
\end{equation} 
and the Ruppeiner metric of the state space may be easily computed from Hessian
matrix of the entropy as
\begin{equation} 
ds^2= \left(\frac{\pi}{q} \sqrt{\frac{p}{q}}- \frac{6}{q^2}\right)dq^2
- \frac{2\pi}{\sqrt{qp}} dqdp
+ \left(\frac{\pi}{p} \sqrt{\frac{q}{p}}- \frac{6}{p^2}\right) dp^2 
\end{equation}
The determinant of the metric is 
$g= -\frac{12}{q^2p^2}(\pi\sqrt{qp}-3)$. 
Hence, the thermodynamic geometry of the state space based on the 
above metric 
is non degenerate in the large charge limit under consideration 
as $g\neq 0$ in this limit. It is now straightforward to compute the 
scalar curvature of the state space to be 
\begin{equation}
R= \frac{\pi \sqrt{qp}}{24\left(\pi\sqrt{qp}-3\right)^3}
\left(\pi^2 qp- 9\pi \sqrt{qp}+18\right) 
\end{equation} 
In the large charge limit for which the asymptotic expansion for the
macroscopic entropy is valid, the scalar curvature is regular as the
numerator is finite and the determinant of the metric $g\neq 0$ in this
limit. It is curious that a viable thermodynamic geometry in the state
space of the two charged extremal black holes being considered here arises
only from the duality invariant macroscopic entropy corrected by the non
holomorphic modifications of the generalised prepotential $F$ of the
special geometry. 

Although typically thermodynamic geometry, particularly Ruppeiner geometry,
is based on the macroscopic thermodynamic entropy an implicit assumption of
a statistical basis in the framework of a canonical ensemble is involved.
Given the mismatch between the duality invariant macroscopic entropy
corrected by non holomorphic modifications of the generalized prepotential
and the microscopic entropy computed in the dual heterotic picture from
degeneracy of appropriate BPS states, it is instructive to explore also the
thermodynamic geometry based on the microscopic entropy expression.  
Although this is obtained from a mixed canonical-microcanonical ensemble in
the OSV picture involving a topological string framework, a comparison of
the scalar curvatures of the thermodynamic geometry based on the two
different entropy expressions may lead to certain physical insight.

The microscopic entropy of the two charged small black
hole from the degeneracy of perturbative BPS heterotic string states is
given as 
\begin{equation} 
S(q,p)= 4 \pi \sqrt{qp}- \frac{27}{4}\ln\left(qp\right) 
\end{equation}
The Ruppeiner
metric for the equilibrium state space of the system based on this entropy
is given by the Hessian matrix of the entropy with respect to
the charges to be
\begin{equation} 
ds^2= \left(\frac{\pi}{q} \sqrt{\frac{p}{q}}- \frac{27}{4q^2}\right)dq^2
- \frac{2\pi}{\sqrt{qp}} dqdp
+\left(\frac{\pi}{p} \sqrt{\frac{q}{p}}- \frac{27}{4p^2}\right) dp^2 
\end{equation}

The determinant of the metric tensor is $ g=
-\frac{27}{16q^2p^2}(8\pi\sqrt{qp}-27) $. Notice that once again the
determinant is non zero for the large charge limit being considered here
and we obtain a non degenerate Ruppeiner geometry even at extremality. The
scalar curvature of the state space in the microscopic picture may now
easily computed to be 
\begin{equation} 
R= \frac{16\pi
\sqrt{qp}}{27\left(8\pi\sqrt{qp}-27\right)^3}
\left(32\pi^2 qp- 324\pi \sqrt{qp}+ 729\right) 
\end{equation}
The scalar curvature is finite and regular in the large charge
limit. The finite scalar curvature indicates as usual a stable
thermodynamic system with an interacting statistical basis. Although the
two scalar curvatures are different in their exact expressions the overall
behavior is similar indicating that the microscopic and the macroscopic
pictures are in close conformity. 

\subsection{Thermodynamic Geometry and Higher $ \alpha^ \prime $ 
Corrections to Small black holes}

\begin{figure}
\centering
\epsfxsize=3.5in
\hspace*{0in}\vspace*{0.2in}
\epsffile{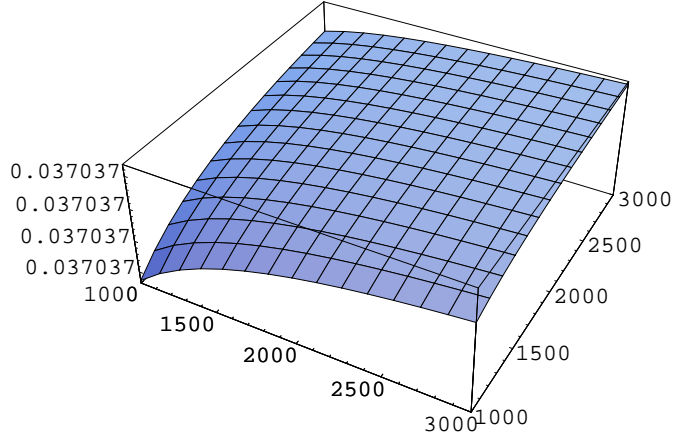}
\caption{\small Ruppeiner curvature for the small black hole with 
$\frac{1}{\sqrt{qp}}$ corrections, as a function of $q$ and $p$}
\label{small1overrootnw}
\end{figure}

It is possible to incorporate further higher derivative corrections to the
microscopic entropy for the two charged small black holes in Type IIA
string theory compactified on $K_3 \times T^2$. It turns out that 
the first subleading order and the logarithmic
corrections are identical to the ones considered before and to the other
subleading orders we have the corrected microscopic entropy as a series 
\begin{equation}
S_{micro} = 4\pi \sqrt{qp} - \frac{27}{4}\ln{qp} + \frac{15}{2}
\ln{2} - \frac{675}{32 \pi \sqrt{qp}} - \frac {6075}{2048 \pi^2 qp}+\cdots
\label{microcorr}
\end{equation}
It is now possible to study the modification to the thermodynamic geometry
arising from the effect of the $\frac{1}{\sqrt{qp}}$ term through the Hessian matrix
of the entropy with respect to the charges. This provides the corrected
Ruppeiner metric as 
\begin{eqnarray} 
ds^2 &=&\left(\frac{\pi}{q} \sqrt{\frac{p}{q}}-
\frac{27}{4q^2}+ \frac{2025}{128\pi q^2\sqrt{qp}}\right)dq^2 -2\left(
\frac{\pi}{\sqrt{qp}}- \frac{675}{128\pi (qp)^{3/2}}\right) dqdp
\nonumber\\ 
&+& \left(\frac{\pi}{p} \sqrt{\frac{q}{p}}- \frac{27}{4p^2}+ 
\frac{2025}{128\pi p^2
\sqrt{qp}}\right) dp^2 
\end{eqnarray} 
The determinant of the metric tensor is 
\begin{equation}
g= -\frac{27}{2048 \pi^2 q^3 p^3} (1024\pi^3 qp \sqrt{qp}- 6656 \pi^2 qp+
16200 \pi \sqrt{qp}- 16875) 
\end{equation}
and is non zero in the large charge limit and we have non
degenerate thermodynamic geometry at extremality.  The scalar curvature of
the state space may now be computed from the Ruppeiner metric. The exact
expression for the scalar curvature is lengthy, and we present it
to the Appendix. In fig. (\ref{small1overrootnw}), we present the 
behavior of the scalar curvature in the large charge limit graphically, as
a function of $q$ and $p$.

It is instructive further to examine whether the other subleading terms
significantly modify the thermodynamic geometry of the equilibrium state
space. To this end we consider next the effect of the $\frac{1}{qp}$
contributions in eq. (\ref{microcorr}) to the entropy on the state 
space geometry and the scalar curvature. The Ruppeiner metric, in this
case, is given by 

\begin{eqnarray}
ds^2 &=& \left(\frac{\pi}{q} \sqrt{\frac{p}{q}}- \frac{27}{4q^2}+ 
\frac{2025}{128\pi q^2\sqrt{qp}}- \frac{6075}{1024\pi^2 q^3p}\right)dq^2
\nonumber\\
&-& 2\left(\frac{\pi}{\sqrt{qp}}- \frac{675}{128\pi (qp)^{3/2}}-
\frac{6075}{2048\pi^2 q^2p^2}\right)dqdp\nonumber\\
&+& \left(\frac{\pi}{p} \sqrt{\frac{q}{p}}- \frac{27}{4p^2}+ 
\frac{2025}{128\pi p^2 \sqrt{qp}}+ \frac{6075}{1024\pi^2qp^3}\right)dp^2
\end{eqnarray}
The determinant of the metric is given as
\begin{eqnarray}
g&=& -\frac{27}{4194304\pi^4q^4p^4} 
\left(2097152\pi^5 p^2q^2\sqrt{qp}+ 30412800\pi^3 pq \sqrt{qp}
\right.\nonumber\\
&-& \left. 13631488\pi^4 q^2p^2- 
22118400\pi^2 pq- 24300000\pi \sqrt{qp}- 4100625\right)
\end{eqnarray}

and is nonzero in the large charge limit and hence defines a non
degenerate thermodynamic geometry at extremality. It is now possible to
compute the thermodynamic scalar curvature and compare with the scalar
curvature of the state space due to corrections upto order $\frac {1}{\sqrt
{qp}}$. The exact expression is provided in the Appendix. In fig.
(\ref{small1overnw}), we present a
graphical analysis of the same. The curvature is non zero and regular in
the large charge limit just as for the other examples that we have
considered. It is observed that the modification to the scalar curvature due to the 
inclusion of the $\frac{1}{qp}$ correction to the entropy is marginal.

\begin{figure}
\centering
\epsfxsize=3.5in
\hspace*{0in}\vspace*{0.2in}
\epsffile{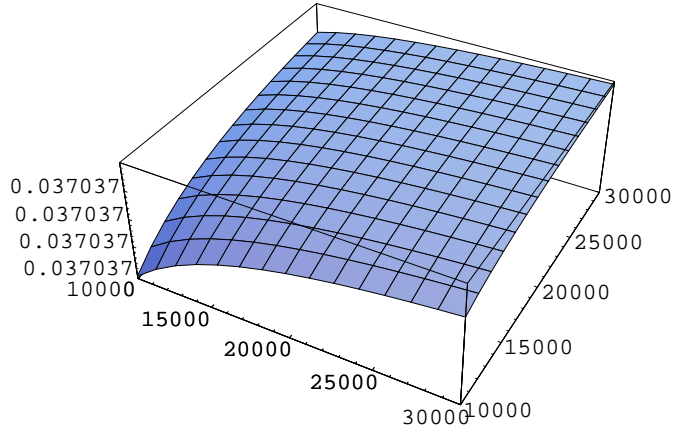}
\caption{\small Ruppeiner curvature with $\frac{1}{qp}$ corrections, as a function
of $q$ and $p$.}
\label{small1overnw}
\end{figure}

\subsection{String Loop Corrections and Exact Entropy Expression.}

Apart from higher derivative terms in the low energy
effective supergravity action there are higher derivative contributions to
the macroscopic entropy also from purely quantum corrections arising from
perturbative string loop corrections and non perturbative contributions.
Hence at the $R^4$ level it is necessary to also consider ${\sqrt
\frac{q}{p}}$ contributions to the macroscopic entropy arising from string
loop corrections. For the two charged small black hole it is possible to
establish the string loop corrected entropy expression from application of
the Sen entropy function formalism and scaling arguments. An explicit
analysis leads to \cite{sinha}
\begin{equation} 
S_{BH} =  \sqrt{aqp + bq}
\end{equation}
with $q \gg p \gg 1$, where $b$ is a constant depending on the loop
corrections and $a$ is an arbitrary constant. Invoking T duality
invariance modifies the entropy expression to
\begin{equation}
 S_{BH} =  \sqrt{aqp + b\left(q+p\right)}
\end{equation}

A modified microscopic state counting for the two charged small black hole
from the degeneracy of BPS states in the dual heterotic string picture
leads to the microscopic entropy $S_{micro} =\sqrt{16qp+\frac{2}{3}q}
$ which is consistent with the $ T$ duality invariant expression for the
macroscopic entropy.  A similar expression from microscopic state counting
also arises from $ D_0 D_4 $ black holes in Type IIA string theory
compactified on $K_3 \times T^2$. These possess near horizon geometry of $
AdS_2 \times S^2 \times CY_3 $ where $ CY_3 \simeq K_3 \times T^2 $.  
Quite surprisingly Linde and Kallosh \cite{linde} has proposed an exact
expression for the entropy from quantum information theory as 
\begin{equation} 
S_{BH} \sim \sqrt{aqp + b(q+p)} 
\end{equation} 
which is also consistent with the T duality invariant expression.

It is instructive to study the thermodynamic geometry of the equilibrium
state space of the two charged small black holes, implied by the exact
entropy expressions. The Ruppeiner metric for the state space from the
Hessian matrix of the exact entropy expression is given as;
\begin{eqnarray}
ds^2&=& \frac{(ap+b)^2}{4\left(aqp+ b(p+q)\right)^{3/2}}dq^2
+\frac{(aq+b)^2}{4\left(aqp+ b(p+q)\right)^{3/2}}dp^2\nonumber\\ 
&+&\left(\frac{(aq+b)(ap+b)}{2\left(aqp+ b(p+q)\right)^{3/2}}- 
\frac{a}{\left(aqp+ b(p+q)\right)^{1/2}}\right)dqdp
\end{eqnarray}
The determinant of the metric is $ g= \frac{ab^2}{4(aqp+ b(p+q))^2}
$ and is non zero everywhere and hence we have a non degenerate
thermodynamic geometry of the equilibrium state space of the two charged
extremal small black hole. The scalar curvature of the state space may now
be computed from the Ruppeiner metric to be
\begin{equation}
R= -\frac{a^2qp+abq+abp+b^2}{b^2 \sqrt{aqp+bq+bp}}.
\end{equation}

Observe that the scalar curvature is non zero but finite once again. As
usual this indicates a stable thermodynamic system with an interacting
microscopic statistical basis. Curiously the scalar curvature of the state
space geometry following from the exact entropy expression is negative. The
significance of the negative curvature is not clear at present.

\section{Summary and Conclusions.}\label{six}

In this paper, we have applied the formalism of thermodynamic geometries to degenerate 
quantum ground states at zero temperatures, exemplified by extremal black holes in 
Type II string theories. Such systems exhibiting macroscopic degeneracies are well 
known in the physics of condensed matter like spin glasses. Our motivation has been
to explore Riemannian geometric structures underlying the equilibrium thermodynamic state 
spaces of extremal black holes in Type II supergravities which arise as low energy limits 
of string theories. As stated, the entropy of extremal black holes suggest a 
limiting characterisation of conventional thermodynamics to degenerate quantum ground 
states at zero temperatures. Our construction is a geometrical realization of the 
equilibrium thermodynamic state space of such systems typified by extremal black holes 
which exhibit macroscopic degeneracy at zero temperatures. 

It is well known that black hole solutions in $N \geq 2$ supergravity involves moduli 
spaces with special K\"ahler geometry. In particular, they exhibit an
attractor phenomena as a consequence of which moduli fields flow under radial evolution 
to fixed values in terms of the
charges at the horizon which is a fixed point of the flow. The entropy is thus a function 
of the charges only, and independent of the asymptotic values of the moduli ensuring 
the validity of an underlying  microscopic statistical basis
in terms of fundamental string states or D-brane systems. The present investigation 
serves as a prelude to explore the thermodynamics of extremal black holes in Type II 
string theories, away from the attractor fixed point and a consequent geometrical 
understanding of the attractor mechanism and the  attractor fixed point as possible 
restrictions in the equilibrium state space extended by the moduli variables.

Quite obviously our construction is a radical departure from the original domain of 
application of the formalism of thermodynamic geometries to conventional thermodynamic 
systems at finite non-zero temperatures. Hence certain
modifications of the scope of the formalism was naturally expected. In particular the 
absence of classical fluctuations
of a thermal origin at zero temperatures rules out the connection between the 
probability distribution of classical fluctuations and a positive definite invariant 
Riemannian form over the equilibrium state space. In fact our construction
clearly illustrates that although a non degenerate thermodynamic geometry emerges, 
the signature of the invariant Riemannian form is indefinite and may also be sensitive 
to the higher derivative corrections to the macroscopic entropy. However, we adopt the 
perspective that given the connection between the scalar curvature and an interacting 
statistical basis for non extremal black holes, the same is valid also for 
extremal black holes. The divergences of the scalar curvature in the case of extremal 
black holes may then possibly describe quantum phase transitions between distinct 
vacua in the moduli space. The attractor mechanism and in particular the 
study of Ferrara et al. \cite{fgk} also suggests a role of the moduli space in 
determining the signature of the invariant interval over the state space and curvature 
singularities with phase transitions amongst vacua in the moduli space.

In the present investigation, we have studied three diverse examples of extremal charged 
black hole solutions of Type II supergravity which arise as low energy limits of 
string compactifications on Calabi-Yau manifolds. These are
four charged black holes in $D=4$ and $D=10$ described by D2-D6-NS5-P brane system, 
three charged black holes in $D=5$ and $D=10$ described by D1-D5-P system and two 
charged small black holes in Type II string theories.
Quite remarkably, we have found that in spite of conventional thermodynamic notions 
being invalid at extremality, the equilibrium state space in all these examples admits of a 
non degenerate Riemannian geometric structure. Although the signature of the invariant 
interval over the state space is no more positive definite. As discussed earlier, this is 
not an issue since classical fluctuations are necessarily absent at zero temperatures. 
We have further computed the scalar curvatures in all these cases and it was found that 
the scalar curvatures were regular everywhere
but vanishingly small in the limit of large charges in which the asymptotic expressions 
for the entropy were valid.
This indicated a possibly weakly interacting statistical basis. The absence of 
divergences in the scalar curvatures implied that the solution were thermodynamically 
stable and no phase transitions were evident. The modification of the geometry and 
the scalar curvature due to higher derivative contributions to the macroscopic entropy 
were also explored. We found that the scalar curvatures were only marginally sensitive 
to the higher derivative modifications of the entropy. One interesting aspect of our 
study was that for the case of the two charged small black holes a non degenerate 
thermodynamic geometry required the inclusion of non holomorphic terms which renders 
the entropy to be duality invariant. The physical significance if any of this issue was 
however not clear from our construction. Furthermore,
we also studied the thermodynamic state space based on an exact entropy expression 
arsing from a quantum information theoretic perspective. The thermodynamic geometry 
based on this entropy was also found to be non degenerate but now with a positive 
definite metric. The scalar curvature was regular but small in the limit of large 
charges and negative. The significance of this is not clear at the present stage of 
our understanding and further investigations in this direction are currently in progress.

\section{Acknowledgments}\label{five} 

We would like to thank Amihay Hanany, Parthasarathi Majumdar, Shiraz
Minwalla, Ashoke Sen and V. Subrahmanyam for discussions. B.N.T. would like to thank CSIR,
India, for financial support under the research grant
CSIR-SRF-9/92(343)/2004-EMR-I, and Vinod Chandra, Ravindra Kumar and Ashoke
Garai for computational help.

\newpage
\begin{center}
{\bf Appendix} 
\end{center}

In this appendix we provide the explicit forms of the various scalar
curvatures for the Ruppeiner geometries of the thermodynamic state spaces of
the examples of extremal charged black holes that we have considered,
including higher derivative corrections.

For the $D=10$ black holes described by the $D2-D6-NS5-P$ systems the scalar
curvature with higher derivative corrections is

\begin{eqnarray}
R &=& \frac{3N_2N_6\sqrt{N_p}N_5^{5/2}}{4 \pi} 
(6C^2\sqrt{N_p}N_2^3N_5^6N_6^3+ 
5CN_2^4N_5^{17/2}N_6^4 \sqrt{N_2 N_5 N_6 N_p} \nonumber \\
&+& 
50C N_2N_5^{5/2}N_6\sqrt{N_2 N_5 N_6 N_p}+ 
N_2^6\sqrt{N_p}N_5^{12}N_6^6+ 100C^4\sqrt{N_p})^{-3} \nonumber \\
& &
(17C^2N_2^{13}N_6^{13}N_5^{53/2}\sqrt{N_2N_5N_6N_p}+
3837C^4N_2^{10}N_6^{10}N_5^{41/2}\sqrt{N_2N_5N_6N_p} \nonumber \\
&+&
565000C^9\sqrt{N_p}N_2^3N_5^6N_6^3-
620000C^{10}N_2N_5^{5/2}N_6\sqrt{N_2N_5N_6N_p}\nonumber \\
&-&
114C^3\sqrt{N_p}N_2^{12}N_5^{24}N_6^{12}+
26C\sqrt{N_p}N_2^{15}N_5^{30}N_6^{15} \nonumber \\
&+&
47472C^6N_2^7N_5^{29/2}N_6^7
\sqrt{N_2N_5N_6N_p}-
800000C^{11}\sqrt{N_p} \nonumber \\
&+&
604500C^8N_2^4N_5^{17/2}N_6^4\sqrt{N_2N_5N_6N_p}+
2N_2^{16}N_5^{65/2}N_6^{16}\sqrt{N_2N_5N_6N_p} \nonumber \\
&+&
20280C\sqrt{N_p}N_2^9N_5^{18}N_6^9+
178980C^7\sqrt{N_p}N_2^6N_5^{12}N_6^6)
\end{eqnarray}
where the constant $C$ has been defined in the main text. 

For the $D=10$ extremal black holes described by D1-D5-P brane system we 
have the scalar curvature of the state space corrected by higher 
derivative $R^4$ contributions arising from the string three loop 
scattering amplitude is given as

\begin{eqnarray}
 R &=&\frac{3}{4} \sqrt{N_1 N_5 N_p} \sqrt{N_1 N_5} (\frac{9 N_1^{10} N_5^{10} b 
\sqrt{N_1 N_5}-256 N_1^7 N_5^7 b^3 \sqrt{N_1 N_5} }{\pi N_p (6 N_1 N_5 \sqrt{N_1 N_5} b^2 
- 20b^3 + N_1^4 N_5^4 \sqrt{N_1 N_5} )^3} \nonumber \\
&-&\frac{- 1920 N_1 N_5 b^7 \sqrt{N_1 N_5}+ 1700 N_1^4 N_5^4 b^5\sqrt{N_1 N_5}+N_1^{12} N_5^{12}
}
{\pi N_p (6 N_1 N_5 \sqrt{N_1 N_5} b^2 - 20b^3
 + N_1^4 N_5^4 \sqrt{N_1 N_5} )^3} \nonumber \\
&+& \frac{6520 N_1^3 N_5^3 b^6-1600 b^8-48N_1^6 N_5^6 b^4 +8 N_1^9 N_5^9 b^2}
{\pi N_p (6 N_1 N_5 \sqrt{N_1 N_5} b^2 - 20b^3
 + N_1^4 N_5^4 \sqrt{N_1 N_5} )^3})
\end{eqnarray}

Next, we provide the Ruppeiner scalar curvature of the state space geometry
based on the microscopic entropy expression for two charged small black
holes in Type II string theory compactified on $K_3 \times T^2$ corrected by
$\frac {1}{{\sqrt qp}}$ contributions :

\begin{eqnarray}
R &=& \frac{1}{8192 n^2 w^2 \sqrt{qp} \pi } [ ( -265420800 \pi ^3 qp \sqrt{qp} - 
29360128 \pi ^5 q^2 p^2 \sqrt{qp} 
\nonumber\\
&+&5948640000 \pi \sqrt{qp}- 7688671875 + 2097152 q^3 p^3 \pi^6 
+ 160350208 q^2 p^2 \pi^4 \nonumber\\
&-& 1149033600 q p  \pi^2 )(1024 \pi^3 qp \sqrt{qp} - 6656 qp \pi^2 
+ 16200 \pi \sqrt{qp} - 16875)^{-3}]\nonumber
\end{eqnarray}

Finally, the Ruppeiner curvature for the two charged small black hole, 
with the $\frac{1}{qp}$ corrections is\\
\begin{eqnarray}
R &=& \frac{65536\pi^3 qp\sqrt{qp}}{27}(2097152\pi^5 p^2q^2\sqrt{qp}+ 
30412800\pi^3 pq \sqrt{qp} \nonumber \\
&-& 13631488\pi^4 q^2p^2- 22118400\pi^2 pq- 
24300000\pi \sqrt{qp}- 4100625)^{-3} \nonumber \\
& & (22823730196875000 \pi \sqrt{qp} +
465892154818560000\pi^5 p^2q^2\sqrt{qp} \nonumber \\
&- &
201080268288000000\pi^3 pq \sqrt{qp}-
192782236070707200\pi^7q^3p^3\sqrt{qp} \nonumber \\
&+& 
3783403212890625- 95786360635392000\pi^9 q^4p^4\sqrt{qp} \nonumber \\
&-&
2744381022928896\pi^{11} q^5p^5\sqrt{qp}-
177903493447680000\pi^6 q^3p^3 \nonumber \\
&+ &
22282702648508416\pi^{10} q^5p^5+
218241931463884800\pi^8 q^4 p^4 \nonumber \\
&+ &
140737488355328\pi^{12} q^6p^6- 9486221850000000\pi^2 pq \nonumber \\
&-&  
42628654080000000\pi^4 q^2p^2) 
\end{eqnarray}

\end{document}